\documentclass[11pt]{article}

\usepackage{amssymb}
\usepackage{amsmath}
\usepackage{amsfonts}
\usepackage{epsfig}


\newcommand{\sect}[1]{\setcounter{equation}{0}\section{#1}}

\newcommand{\Tr}{{\,\rm Tr}\:}
\newtheorem{theorem}{Theorem}[section]

\newtheorem{remark}{Remark}[section]
\newtheorem{proposition}{Proposition}[section]
\newtheorem{lemma}{Lemma}[section]
\newtheorem{corollary}{Corollary}[section]
\newtheorem{definition}{Definition}[section]
\def\br{\begin{remark}\rm\small}
\def\er{\end{remark}}
\def\bt{\begin{theorem}}
\def\et{\end{theorem}}
\def\bd{\begin{definition}}
\def\ed{\end{definition}}
\def\bp{\begin{proposition}}
\def\ep{\end{proposition}}
\def\bl{\begin{lemma}}
\def\el{\end{lemma}}
\def\bc{\begin{corollary}}
\def\ec{\end{corollary}}
\def\beaq{\begin{eqnarray}}
\def\eeaq{\end{eqnarray}}

\newcommand{\Res}{\mathop{\,\rm Res\,}}

\begin{document}

\setlength{\baselineskip}{5.0mm}

\pagestyle{empty}
\hfill CERN-PH-TH-2009-233

\addtolength{\baselineskip}{0.20\baselineskip}
\begin{center}
\vspace{26pt}
{\large \bf {Chain of matrices, loop equations and topological recursion }}
\newline
\vspace{26pt}

\noindent
{{\sc Nicolas Orantin}$^1$,\\
Theory Division,
CERN,\\
Geneva 23, CH-1211 Switzerland\\
}

\end{center}

\begin{center}
{\bf Abstract}
\end{center}
Random matrices are used in fields as different as the study of multi-orthogonal polynomials or the enumeration of
discrete surfaces. Both of them are based on the study of a matrix integral. However, this term can be confusing
since the definition of a matrix integral in these two applications is not the same.
These two definitions, perturbative and non-perturbative, are discussed in this chapter as well
as their relation. The so-called loop equations satisfied by integrals over random matrices coupled in chain
is discussed as well as their recursive solution in the perturbative case when the matrices are Hermitean.

\vspace{1cm}

\sect{Introduction: what is a matrix integral?}\label{intro}

The diversity of aspects of mathematics and physics exposed in the present volume witnesses how rich the theory of
random matrices can be. This large spectrum of applications of random matrices does not only come from the numerous
possible ways to solve it but it is also intrinsically due to the existence, and use, of different definitions of the matrix
integral giving rise to the partition function of the theory under study.

Back to the original work of Dyson \cite{Dyson}, the study of random matrices is aimed at computing integrals over some
given set of matrices with respect to some probability measure on this set of matrices. In order to be computed, these
integrals are obviously expected to be convergent. Nevertheless, one of the main applications of random matrices
in modern physics follows from a slightly different definition. Following the work of \cite{BIPZ}, the matrix integral can be considered,
through its expansion around a saddle point of the integrand, as a formal power series seen as the generating function
of random maps, i.e. random surfaces composed of polygons glued by their sides\footnote{See chapter 26 for
an introduction this topic and \cite{eynform} and reference therein for the generalization to multi-matrix
integrals.}. Whether this formal series has a non-vanishing
radius of convergency or not does not make any difference: only its coefficients, which take finite values, are meaningful.

The issue whether these two definitions do coincide or not was not addressed for a long time and led to confusions. In
particular it led to a puzzling non-coincidence of some result in the literature \cite{AkAm,BrezinDeo,Kanzieper}. Their computations of the same quantity, even if proved to be right, did not match. This puzzled was
solved by Bonnet, David and Eynard \cite{BDE} who were able to show that the mismatch between the two solutions is a consequence
of the discrepancy between the definitions of the matrix integrals taken as partition functions.

Since some of the topics discussed in the present chapter do depend on the definition one considers for the partition function whereas some other issues do not, section \ref{secconv} is devoted to the precise definition of these different matrix integrals. In section \ref{secloop}, we present the loop equations which can be used to compute the
partition function and correlation functions of a large family of matrix models.
Section \ref{sec1MM} is devoted to a review of one the solution of the one Hermitean matrix model through the use of
the so-called loop equations. Section \ref{secChain} generalizes this method to an arbitrary number of Hermitean
matrices coupled in chain. Finally, section \ref{conclusion} gives a short overview of generalizations and applications
of this very universal method.

\sect{Convergent vs formal matrix integral} \label{secconv}

One of the most interesting features in the study of random matrices is the behavior of the statistic of eigenvalues, or
correlation functions, as the random matrices become arbitrary large. This limit is not only very interesting for its
applications in physics (study of heavy nuclei, condensed matter...)  but also in mathematics: the knowledge of the large
size limit allows to access the asymptotics of a large set of multi-orthogonal polynomials\footnote{See \cite{Mehta} for a nice review of these application and all the other chapters of the present volume.}.

Most of the usual technics used in random matrix theory fail in the study of the large
matrix limit.
However, one possible way to address this problem is to try to use naively some saddle point analysis.
Let us consider the example of a Hermitean one matrix with polynomial potential to illustrate this procedure. The partition
function is given by the matrix integral:
$$
{\cal Z}(V) = \int_{{\cal H}_N} dM e^{-{N \over t} \Tr V(M)}
$$
where one integrates over the group ${\cal H}_N$ of Hermitean matrices of size $N$ with respect to the
measure
$$
dM:= {{\displaystyle \prod_{i=1}^N} k! \over \pi ^{N(N-1)\over 2}}\prod_{i=1}^N dM_{ii} \prod_{i<j} d\Re\left(M_{ij}\right) \, d \Im\left(M_{ij}\right)
$$
defined as the product of the Lebesgues mesures of the real components of the matrix $M$ divided by the volume of the
unitary group of size $N$.
For the sake of simplicity,
one assumes that the potential $V(x)= {\displaystyle \sum_{k=0}^{d}} {t_k \over k+1} x^{k+1}$ is a polynomial. {\em Notice that the direct saddle
point analysis of this integral does not make sense in general.}

In order to fix this, let us consider a more general problem. Instead of considering Hermitean matrices, we
consider normal matrices of size $N$ whose eigenvalues lie on some arbitrary path $\gamma$ in the complex plane:
$H_N(\gamma)$ is the set of matrices $M$ of size $N \times N$ such that there exists $U \in U(N)$ and $X = \hbox{diag}(x_1, \dots, x_N)$
with $x_i \in \gamma$ satisfying $M = U X U^\dagger$.

With this notation, the set of Hermitean matrices is $H_N\left(\mathbb{R}\right)$. Given a fixed potential $V(x)$,
one considers the family of matrix integrals over formal matrices on arbitrary contours $\gamma$:
$$
{\cal Z}(V,\gamma) = \int_{{\cal H}_N(\gamma)} dM e^{-{N \over t} \Tr V(M)}.
$$
As in the Hermitean case, one can integrate out the unitary group to turn this partition function into an integral over
the eigenvalues of the random matrix\footnote{This procedure can be generalized to multi-matrix models using the
HCIZ formula \cite{IZ,HC} presented in chapter 17.}:
$$
{\cal Z}(V,\gamma) = \int_{\gamma} \dots \int_{\gamma} \, {\displaystyle \prod_{i=1}^N} dx_i \, \prod_{i<j} (x_i-x_j)^2 \,
e^{-{N \over t} {\displaystyle \sum_{i=1}^N} V(x_i)}.
$$
However, given a polynomial potential of degree $d+1$, not every path $\gamma$ is admissible. Indeed, there are only
$d+1$ directions going to infinity where $\Re\left[V(x)\right]>0$ as $x\to \infty$ and where the integrand decreases rapidly enough
for the integral to converge. Thus there exists $d$ homologically independent paths on which the integral $\int dx e^{-{N \over t} V(x)}$
is convergent. Let us choose a basis $\{\gamma_i\}_{i=1}^d$ of such paths. Every admissible path $\gamma$ for the eigenvalues of
the random matrix can thus be decomposed in this basis:
$
\gamma = {\displaystyle \sum_{i=1}^d} \kappa_i \gamma_i.
$

Using this decomposition, for any admissible path $\gamma$, the partition function reduces to
\begin{equation}\label{decompoZ}
{\cal Z}(V,\left\{\gamma_i\right\}|\left\{\kappa_i\right\}) = N! \, \sum_{\{n_i\}} {\displaystyle \prod_{i=1}^d} { \kappa_i^{n_i} \over  n_i!}
\int_{\gamma_1^{n_1}\times \dots \times \gamma_d^{n_d}} \, {\displaystyle \prod_{i=1}^N} dx_i \, \prod_{i<j} (x_i-x_j)^2 \,
e^{-{N \over t} {\displaystyle \sum_{i=1}^N} V(x_i)}
\end{equation}
where one sums over all integer $d$-partitions $(n_1,\dots,n_d)$ of $N$, i.e. the sets of $d$ integers
$\{n_i\}_{i=1}^d$ satisfying $n_1+ \dots + n_d = N$.


The requirement of convergence of the integral only fixes the asymptotic directions of the paths $\gamma_i$'s.
We still have the freedom to choose their behavior away from their asymptotic directions.
Does there exist one choice better than the others?
One is interested in performing a saddle point analysis of the matrix integral. One thus has to look for the singular points
of the action, i.e. the solutions of $V'(x) = 0$. There exist $d$ such solutions $\xi_i$, $i=1,\dots,d$, i.e. as many
as the number of paths $\gamma_i$ in one basis. In the case
of the one matrix model with polynomial potential exposed in the present section, it was proved following \cite{bertolaboutroux} that
there exists a good basis in the sense that every path $\gamma_i$ is a steepest descent contour\footnote{The existence of
a good path is conjectured to hold for all other matrix models discussed in this chapter. However, the proof is known, at the
time these lines are written, only in the one matrix model case.}. More precisely, along any path $\gamma_i$, the effective
potential felt by an eigenvalue $x$, $V_{eff}(x) = V(x) - {t \over N} \left< \ln \left( \det x-M \right) \right>$, behaves
as follows: its real part decreases then stays constant on some interval and then increases, whereas its imaginary part
is constant then increasing and finally constant.

Such a steepest descent path can thus be seen as a possible vacuum for one eigenvalue. Each $d$-partition
of $N$ hence corresponds to one vacuum for the theory, or one saddle configuration for the random matrix.
The formula eq.\ref{decompoZ} can be understood as a sum over all possible vacua of the theory:
$$
{\cal Z}(V,\gamma) = \sum_{n_1+ \dots + n_d = N} N! {\displaystyle \prod_{i=1}^d} { \kappa_i^{n_i} \over  n_i!}
{\cal Z}(V,\left\{\gamma_i\right\}|n_1,\dots,n_d)
$$
where the partition function with fixed filling fractions $\epsilon_i = {n_i \over N}$
$$
{\cal Z}(V,\left\{\gamma_i\right\}|n_1,\dots,n_d):=\int_{\gamma_1^{n_1}\times \dots \times \gamma_d^{n_d}} \, {\displaystyle \prod_{i=1}^N} dx_i \, \prod_{i<j} (x_i-x_j)^2 \,
e^{-{N \over t} {\displaystyle \sum_{i=1}^N} V(x_i)}
$$
is the weight of a fixed configuration of eigenvalues, or the partition function of the theory with a fixed vacuum labeled
by a partition $(n_1, \dots, n_d)$.

{\em Assuming that the paths $\gamma_i$ are good steepest descent paths}, the partition functions with fixed filling fractions
can be computed by saddle point approximation, i.e. perturbative expansion of the integral around a saddle as $t \to 0$. Further {\em assuming that one can commute the integral and the power series expansion},
the result is a formal power series in $t$ whose coefficients are gaussian matrix integrals:
$$
{\cal Z}(V,\left\{\gamma_i\right\}|n_1,\dots,n_d) \sim {\cal Z}_{formal}(V,\left\{\gamma_i\right\}|n_1,\dots,n_d) \qquad \hbox{when} \qquad t \to 0
$$
with
\begin{eqnarray}
{\cal Z}_{formal}&:=& e^{-{N \over t} \sum_i n_i V(\xi_i)} {\displaystyle \sum_{k=0}^\infty}
{(-1)^k N^k \over t^k \, k!}  {\displaystyle \prod_{i=1}^d}
\left(\int_{H_{n_i}(\gamma_i)} dM_i\right) \left( {\displaystyle \sum_{i=1}^d} \Tr \delta V_i (M_i)\right)^k\cr
&& \times e^{- {N \over 2t} {\displaystyle \sum_i} V''(\xi_i) \Tr (M_i - \xi_i {\mathbf{1}}_{n_i})^2}
{\displaystyle \prod_{j<i}} \det \left( M_i \otimes {\mathbf{1}}_{n_j} - {\mathbf{1}}_{n_i} \otimes M_j\right)^2 \cr
\end{eqnarray}
where $\left\{\xi_i\right\}_{i=1..d}$ denote the $d$ solutions of the saddle point equation $V'(\xi_i) = 0$,
$\delta V_i(x) := V(x) - V(\xi_i) - {V''(\xi_i) \over 2} (x - \xi_i)^2$ denotes the non-gaussian part of the Taylor
expansion of the potential around the saddle $\xi_i$ and the notation ${\displaystyle \prod_{i=1}^d}
\left(\int_{H_{n_i}(\gamma_i)} dM_i\right)$ stands for the multiple integral $\int_{H_{n_1}(\gamma_1)} dM_1 \dots \int_{H_{n_d}(\gamma_d)} dM_d$.
This formal series in $t$ is referred to as a {\em formal matrix integral} even though {\bf it is not a matrix integral} but a formal power series in $t$.

This construction can be thought of as a perturbation theory: the matrix integral ${\cal Z}(V)$ is the non-perturbative
partition function of the theory whereas the formal matrix integral ${\cal Z}_{formal}(V,\mathbb{R}|n_1,\dots,n_d)$
is a perturbative partition function corresponding to the expansion around a fixed vacuum $(n_1,\dots,n_d)$ in the
basis $(\gamma_1,\dots,\gamma_d)$.

Since these two possible definitions of the partition function might be confused, let us emphasize their main differences,
concerning their properties as well as their applications:
\begin{itemize}

\item The convergent matrix integral is fixed by a choice of potential $V$ together with an admissible path
$\gamma$. The formal matrix integral depends on a potential $V$ of degree $d$, a basis of admissible paths $\left\{\gamma_i\right\}_{i=1}^d$ and a $d$ partition of $N$, $\left\{n_i\right\}_{i=1}^d$.

\item By definition, the non-perturbative partition function is a convergent matrix integral for arbitrary potential,
provided the paths $\gamma_i$ are chosen consistently. The perturbative integral is a power series defined for arbitrary
potentials, integration paths and filling fractions. It might be non-convergent, and will be for most of combinatorial
applications;

\item The logarithm of the perturbative partition function always has a ${1 \over N^2}$ expansion, whereas the
non-perturative one does not have one most of time (see section \ref{topoexpansion}).

\item  The formal matrix integral is typically used to solve problems of enumerative geometry such as enumeration of
maps or topological string theory. The convergent matrix integral is related
for example to the study of multi-orthogonal polynomials.

\end{itemize}

\sect{Loop equations} \label{secloop}

Even if the perturbative and non-perturbative partition functions do not coincide in general, they share some common
properties. One of the most useful is the existence of a set of equations linking the correlation functions of the
theory: the loop equations. These equations, introduced by Migdal \cite{Migdalloop}, are simply the Schwinger-Dyson equations
applied to the matrix model setup. They proved to be an efficient tool for the computation of formal matrix
integrals as the explicit computation of one class of one Hermitean formal matrix integral  by Ambjorn and al \cite{ACKM} proves.

\subsection{Free energy and correlation functions}
One of the main quantities used in the study of matrix integrals is the {\em free energy} which is defined as the
logarithm of the partition function:
$$
{\cal F} := - {1 \over N^2} {\cal Z}.
$$
In the formal case, where ${\cal Z}$ is the generating function of closed discrete surfaces, the free energy
enumerates only connected such surfaces.

In order to be able to compute the free energy, but also for their own interpretation in combinatorics of maps or
string theory\footnote{They are generating functions of open surfaces, as opposed to the free energy which generates
surfaces without boundaries. The interested reader can refer to chapter 31 of the present book or the review
\cite{eynform} for details of this interpretation.}, it is convenient to introduce the following correlation functions:
$$
W_k(x_1,\dots,x_k) := \left<\Tr {1 \over x_1-M} \Tr {1 \over x_2-M} \dots \Tr {1 \over x_k-M} \right>_{c}
$$
where the index $c$ denotes the connected part and
\begin{equation}\label{defreso}
{1 \over x-M} = {\displaystyle \sum_{i=1}^d \sum_{k=0}^\infty} {\left(M_i-\xi_i \mathbb{I}_{n_i}\right)^k \over (x-\xi_i)^{k+1}}
\end{equation}
 has to be
understood as a formal power series. It is also useful to introduce the polynomial of degree $d-1$ in $x$
$$
P_k(x,x_1,\dots,x_k) := \left<\Tr {V'(x) -V'(M) \over x-M} {\displaystyle \prod_{i=1}^k} \Tr {1 \over x_i-M} \right>_{c}.
$$

\subsection{Loop equations}

The non-perturbative partition function is given by a convergent matrix integral. It should thus be invariant under
change of the integration variable $M$ (or its entries). The name loop equation refers to any equation obtained from the invariance
to first order in $\epsilon \to 0$ of the partition function under a change of variable of the form
$M \to M + \epsilon \delta(M)$\footnote{It can be equivalently seen as integration by parts.}:
$$
\int_{{\cal H}_N(\gamma)} dM e^{-{N \over t} \Tr V(M)} =
\int_{{\cal H}_N(\gamma)} d(M+ \epsilon \delta(M)) e^{-{N \over t} \Tr V(M+ \epsilon \delta(M))}.
$$
To first order in $\epsilon$, this means that the variation of the action should be compensated by the Jacobian
of the change of variables:
$$
{N \over t} \left< \Tr V'(M) \delta(M)\right> = \left< J(M) \right>.
$$
Actually, the form of the changes of variable considered is limited to two main families of $\delta(M)$. This allows to
give a recipe to compute the Jacobian rather easily as follows.
\begin{itemize}

\item Leibnitz rule:
$$J\left[A(M)B(M)\right] = \left\{J\left[A(M)B(m)\right]\right\}_{m \to M} + \left\{J\left[A(m)B(M)\right]\right\}_{m \to M};$$

\item Split rule:
$
J\left[A(m)\, M^l \, B(m)\right] = {\displaystyle \sum_{j=0}^{l-1}} \Tr\left[A(m) M^j\right] \Tr\left[ M^{l-j-1} B(m)\right];
$

\item Merge rule:
$
J\left[A(m) \Tr \left(M^l B(m) \right) \right] = {\displaystyle \sum_{j=0}^{l-1}} \Tr\left[A(m) M^j  B(m) M^{l-j-1}\right];
$

\item if there is no $M$: $J\left[A(m)\right] = 0$.

\end{itemize}

The formal matrix integral is obtained from Gaussian convergent integrals by algebraic computations which commute with
the loop equations thus
\bt
The formal matrix integrals satisfy the same loop equations as the convergent matrix integrals.
\et

\subsection{Topological expansion}\label{topoexpansion}

The loop equations are a wonderful tool for the study of formal matrix integrals. From now on, we restrict our study
to these formal power series, leaving aside the convergent matrix integrals.

Following an observation originally made by t'Hooft in the study of Feynman graphs of QCD \cite{thooft}, one can
see that the exponent of $N$ in the free energy ${\cal F}$ is the Euler characteristic of the surface enumerated
by this partition function. Thus, ${\cal F}$ admits a ${1 \over N^2}$ expansion
$$
{\cal F} = \sum_{g=0}^\infty N^{-2g} F^{(g)}
$$
commonly called {\em topological expansion} since the terms $F^{(g)}$ of this expansion are generating functions of
connected closed surfaces of fixed genus $g$.

As for the free energy, one can collect  together coefficients with the same power of $N$ in the correation
function and get
$$
W_k(x_1,\dots,x_k) = \sum_{h=0}^\infty \left({N \over t}\right)^{2-2h-k} W_k^{(h)}(x_1,\dots,x_k)
$$
as well as
$$
P_k(x,x_1,\dots,x_k) = \sum_{h=0}^\infty \left({N \over t}\right)^{1-2h-k}
P_k^{(h)}(x,x_1,\dots,x_k)
$$
where the coefficient are formal power series in $t$ independent of $N$.

Both the convergent (non-perturbative) partition function and the formal (perturbative) matrix integral are solution to
the loop equations. Nevertheless they do not coincide in general considered that the loop equations have not a unique solution.
Indeed, in order to make the solution of these equations unique, one has to fix some "initial conditions" satisfied by
the sought for solution and the convergent and formal matrix integrals are not constrained by the same kind of conditions.

On the one hand, the formal matrix integral has well defined constraints: it has a ${1 \over N^2}$ expansion and the
small $t$, large $x$, limit of any correlation function is fixed by the choice of filling fractions. In other words, by
fixing the filling fractions, one prevents the eigenvalues of the random matrix from tunneling from one saddle to
another, i.e. from one steepest descent path to another. There is no instanton contribution.

On the other hand, the convergent matrix integral does not admit, in general, any ${1 \over N^2}$ expansion. Moreover,
its resolvent is not normalized by any arbitrarily fixed choice of filling fraction: it is rather normalized by
some equilibrium conditions on the configuration of the eigenvalues, which, thanks to tunneling, gives instanton corrections
to the classical partition function around the true vacuum of the theory. This means that the eigenvalues of the
matrix distribute on the different paths of the basis in such a way that they are in equilibrium under the action of
the potential and their mutual logarithmic repulsion.

In the formal case, one of the main properties of the correlations functions is the existence of a topological
expansion. Let us plug these topological expansions into one set of equations obtained by considering the change of variable of type
$\delta M = {1 \over x-M} {\displaystyle \prod_{i=1}^k} \Tr {1 \over x_i-M}$. They read, order by order in $N^{-2}$:
\begin{equation}\label{loopeq}
\begin{array}{rcl}
V'(x) W_{n+1}^{(h)}(x,J) &=&  W_{n+1}^{(h-1)}(x,x,J) + {\displaystyle \sum_{m=0}^h \sum_{I \subset J}} W_{1+\left|I\right|}^{(m)}(x,I) W_{1+n-\left|I\right|}^{(h-m)}(x,J\backslash I) \cr
&& + P_{n}^{(h)}(x,J)
+ {\displaystyle \sum_{i=1}^n } {\partial \over \partial x_j} {W_{n}^{(h)}(x,J\backslash\{x_j\}) - W_n^{(h)}(J) \over x-x_j}\cr
\end{array}
\end{equation}
where $J$ stands for $\left\{x_1,\dots,x_n\right\}$.
This is the hierarchy of equations which is solved in the following section.

\br
Remember that the correlation functions can be seen as the generating functions of discrete surfaces of given topology. In this picture, the loop equations get a combinatorial interpretation: they summarize all the possible
ways of erasing one edge from surfaces of a given topology. This gives a recursive relation among generating
functions of surfaces with different Euler characteristics. This inductive method was introduced in the case of
triangulated surface by Tutte \cite{tutte} without any matrix model's representation of the considered generating
functions.

\er


\sect{Solution of the loop equations in the 1MM}\label{sec1MM}

The solution of the loop equations in their topological expansion has been under intensive study since their
introduction by Migdal \cite{Migdalloop}. In particular, \cite{ACKM} proposed a general solution of these equations in the
one matrix model case for the so-called one cut case, i.e. the case where only one of the filling fractions
$\epsilon_i$ doesn't vanish. The first steps in the study of the 2-cut case were then performed by Akemann in
\cite{Ak96}.

Later, in 2004, Eynard \cite{eynloop1mat} solved the loop equations eq.\ref{loopeq} for  the formal integral
for an arbitrary number of cuts, i.e. compute all the
terms in the topological expansion of any correlation function as well as the free energy's ${1 \over N^2}$-expansion for arbitrary filling fractions.
This solution relies heavily on the existence of an algebraic curve encoding all the properties of the considered matrix
model: the spectral curve. Let us first remind how the latter can be derived.

\subsection{Spectral curve}

Consider eq.\ref{loopeq} for $(h,n) = (0,0)$: it
is a quadratic equation satisfied by the genus 0 one point function:
\begin{equation}\label{masterloop}
W_1^{(0)}(x)^2 -V'(x) W_1^{(0)}(x) = - P^{(0)}(x)
\end{equation}
called the master loop equation. This can be written as an algebraic equation
$$
H_{1MM}(x,W_1^{(0)}) = 0
$$
where $H_{1MM}(x,y)$ is a polynomial of degree $d$ in $x$ and 2 in $y$.
The algebraic
equation $H_{1MM}(x,y)=0$ is the basis of the solution presented in this section and will be referred to
as the {\em spectral curve} of the considered matrix model.

A first corollary of this equation is the multi-valuedness of  $W_1^{(0)}(x)$ as a function of $x$. Indeed,
considered $P(x)$ known, one can solve this equation and get:
\begin{equation}\label{W10}
W_1^{(0)}(x)= {V'(x) \pm \sqrt{V'(x)^2 - 4 P^{(0)}(x)} \over 2}.
\end{equation}
A priori, for any value of the complex variable $x$, there exist two values of $W_1^{(0)}(x)$. Since, from the
definition \ref{W10}, its large $x$ behavior is known to be
\begin{equation}\label{prop1}
W_1^{(0)}(x) \sim {t \over x} {{\displaystyle \sum_{i=1}^d} n_i \over N} = {t \over x} \; \;  \hbox{as} \; \;  x \to \infty,
\end{equation}
one has to select the $-$ sign in order to get the physically meaningful correlation function.

If one can relieve this ambiguity by hand for the genus zero one point function, the computation of the complete
topological expansion of all the correlation functions would imply such a choice at each step.

On the other hand, one can totally get rid of this problem by understanding where it originates from. Any correlation
function is defined as a formal power series both in $t \to 0$ and in $x \to \infty$. The coefficients of the
${1 \over N^2}$-expansion of $W_1(x)$ are thus well defined only around $x \to \infty$, as this series might have a finite
radius of convergency: it is not an analytic function of $x$. In order to get a monovalued function, one has to extend
this series further than this radius. The master loop equation tells us how one can proceed: instead of considering
the correlation function $W_1(x)$ as a function of the complex variable $x$, one should consider it as a function
defined on the spectral curve. That is to say that one should not consider $W_1^{(h)}(x)$ as functions
of a complex variable $x$ but rather as functions of a complex variable $x$ together with a $+$ or $-$ sign
corresponding to the choice of one branch of solution of the master loop equation. The tools of algebraic geometry are built to be able to deal with such situations and we present it in the next section.

\subsection{Algebraic geometry}

Consider an algebraic equation ${\cal E}(x,y) = 0$ in $y$ and $x$ of respective degrees degree $d_y+1$ and $d_x+1$.

A classical result of algebraic geometry states that there exists a compact Riemann surface ${\cal L}$ and two meromorphic
functions $x(p)$ and $y(p)$ on it such that:
$$
\forall p \in {\cal L} \, , \; {\cal E}(x(p),y(p)) = 0.
$$
By abuse of language, one shall use the term spectral curve to denote the Riemann surface ${\cal L}$, the triple
$({\cal L},x,y)$ and the equation ${\cal E}(x,y) = 0$ in the following, when no ambiguity can occur.

Let us detail some general properties of this spectral curve useful for the resolution of the matrix model\footnote{
Most of the properties needed for the study of matrix models can be found in \cite{Farkas,Fay} as well chapter 29
of this volume.}.

\subsubsection{Sheeted structure and branch points}

For a generic fixed value of $x$, there exists $d_y+1$ functions of $x$, $y^i(x)$, $i=0,\dots,d_y$ solution of the equation
${\cal E}(x,y^i(x)) = 0$. In other words, a given complex number $x(p)$ has $d_y+1$ preimages $p^i$, $i=0,\dots,d_y$
on the surface ${\cal L}$ corresponding to different values of $y(p^i)$: one can thus see the Riemann surface ${\cal L}$
as $d_y+1$ copies of the Riemann sphere, denoted as $x$-sheets, glued together, the function $x$ being injective on each copy\footnote{Each copy of
the Riemann sphere corresponds to a branch of solution in $y$ of the equation ${\cal E}(x,y)=0$}.

How are these sheets glued together to form the Riemann surface ${\cal E}$? Two sheets merge when two branches of solution
in $y$ coincide: $y(p^i) = y(p^j)$ for $i \neq j$. These critical points $a_i$, called {\em branch points}, are characterized by
the vanishing of the differential $dx$, i.e. the branch points are solutions of the equation $dx(a_i) = 0$. From now on,
we suppose that all the branch points are simple zeroes of the one form $dx$. This means that only two sheets merge at these
points.

This last assumption implies that, around a branch point $a$, one has
$$
y(p) \sim y(a) + \sqrt{x(p) - x(a)}.
$$
This assumption also implies that, for any branch point $a_i$ and any point $z$ close to $a_i$, there exists a unique point
$\overline{z}$ such that $x(\overline{z}) = x(z)$ and $\overline{z}\to a_i$ as $z \to a_i$\footnote{The application $z \to \overline{z}$
is defined only locally around the branch points and depends on the branch point considered and the notation $\overline{z}$
is abusive. Nevertheless, this application will always be used in the vicinity of a branch point in such a way that no
ambiguity will occur.}.
We call $\overline{z}$ the point conjugated to $z$ around $a_i$.

The spectral curve ${\cal L}$ is thus a $d_y+1$ covering of the Riemann sphere with simple ramification points solutions
of $dx(a_i)=0$.

\smallskip

{\bf Example: hyperelliptic curve}

Let us consider an hyperelliptic spectral curve, i.e. a curve given by a quadratic equation $d_y=1$, as in the one Hermitean matrix model case:
$$
{\cal{E}}(x,y) = y^2 - \prod_{i = 1}^{d}  (x-x(a_i)) (x-x(b_i)) = y^2 - \sigma(x)
$$
where $d_x:= 2 d$ to match the notations of the previous section.
The corresponding Riemann surface can be seen as a two sheeted cover of the Riemann sphere:
 one sheet corresponding to the
branch $y(x) = \sqrt{\sigma(x)}$, and the other one to the other branch $y(x) = - \sqrt{\sigma(x)}$.
These two sheets merge when $y(x)$ takes the same value on both sheets, i.e. when $y(x)$ vanishes. The branch points are
thus the preimages of the points $x(a_i)$ and $x(b_i)$ on the spectral curve. The latter can thus be described as two copies
of $\mathbb{CP}^1$ glued by $d$ cuts $[a_i,b_i]$.

\subsubsection{Genus and cycles}

Generically, the compact Riemann surface ${\cal L}$ associated to an algebraic equation may have non vanishing genus $g$,
and it will be the case in most of the applications of the present chapter.

The Riemann-Hurwitz theorem allows us to get this genus out of the branched covering picture of the spectral curve. For example,
if there are only simple ramification points, it
states that
$$
g = - d_y + {\hbox{number of branch points} \over 2}
$$
where $d_y+1$ is the number of $x$-sheets.

If the Riemann surface has non-vanishing genus, i.e. it is not conformally equivalent to $\mathbb{CP}^1$, there exist
non-contractible cycles on it. In order to deal with them, it will be useful to choose a canonical homology basis
of cycles $\left({\cal A}_1,\dots,{\cal A}_g,{\cal B}_1,\dots,{\cal B}_g\right)$ satisfying the intersection conditions
$$
\forall i,j = 1 \dots, g \, , \; {\cal A}_i \bigcap {\cal A}_j = {\cal B}_i \bigcap {\cal B}_j = 0
\;\; \; , \; \; \; {\cal A}_i \bigcap {\cal B}_j = \delta_{i,j}.
$$

\smallskip

{\bf Example: hyperelliptic curve}

Let us keep on considering the example of an hperelliptic curve with $2d$ branch points. From the Riemann-Hurwitz theorem,
it has genus $g = d-1$. This follows the intuitive picture of two Riemann sphere glued by $d$ segments
giving rise to a genus $d-1$ surface.

\begin{figure}[h]
\unitlength1cm
\begin{center}
\epsfig{file=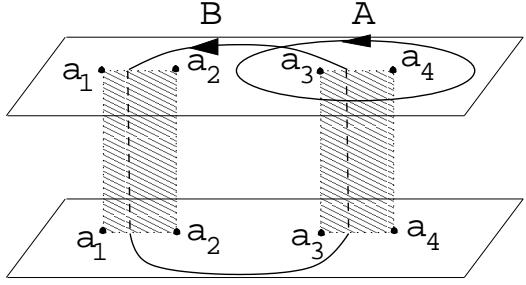,height=4cm}
\end{center}
\caption{Genus 1 hyperelliptic curve: it is built as two copies of the Riemann sphere glued by two cuts
$[a_1,a_2]$ and $[a_3,a_4]$. The unique ${\cal A}$-cycle encircles $[a_3,a_4]$ while the ${\cal B}$-cycles goes
through both cuts.
\label{figcycles}}
\end{figure}

One can also explicit a canonical homology basis as follows. First choose one cut, for example $[a_1,b_1]$ and one sheet,
e.g. the sheet corresponding to the minus sign of the square root.
Then define the ${\cal A}_i$-cycle as the cycles on the chosen sheet around the cut $[a_{i+1},b_{i+1}]$. Finally,
define the ${\cal B}_i$-cycle as the composition of the segment $[a_1,a_{i+1}]$ in the chosen sheet and $[a_{i+1},a_1]$
followed in the opposite direction in the other sheet(see fig.\ref{figcycles} for the simplest example of a genus 1 surface).

\subsubsection{Differentials}

The meromorphic differentials on the Riemann surface ${\cal L}$ and their properties will play a crucial role in the following.
In particular, let us remind a fundamental result concerning meromorphic differentials: a meromorphic differential $df$
on Riemann surface ${\cal L}$ of genus $g$ equipped with a basis of cycles $\left\{{\cal A}_i,{\cal B}_i\right\}_{i=1}^g$,is
defined uniquely by its ${\cal A}$-cycles $\int_{{\cal A}_i} df$ on the one hand
and its singular behavior, i.e. the position of its poles and the divergent part of its Laurent expansion around the latter.



For example, one introduce one of the main character of the resolution of loop equations as follows:
\bd
Let the Bergman kernel $B(p,q)$ be the unique bi-differential in $p$ and $q$ on ${\cal L}$ defined by the
constraints as a differential in $p$:
\begin{itemize}
\item  it has a unique pole located at $p \to q$ which is double without residue. In local coordinates, it reads
$$
B(p,q) \sim {dp \, dq \over (p-q)^2} + \hbox{regular} \qquad \hbox{when} \qquad p \to q ;
$$

\item it has vanishing ${\cal A}$-cycle integrals:
$$
\forall i = 1,\dots, g \, , \; \oint_{{\cal A}_i} B(p,q) = 0 .
$$

\end{itemize}

\ed

It is also useful to define the primitive of the Bergman kernel:
$$
dS_{p_1,p_2}(q) = \int_{z=p_1}^{p_2} B(z,q)
$$
which is a one form in $q$ with simple poles in $q \to p_1$ and $q \to p_2$ with respective residues $-1$ and $+1$.

\subsection{The one point function and the spectral curve}

With these few elements of algebraic geometry in hand, let us complete our study of the spectral curve of the Hermitean
one matrix model. Up to now, one has obtained that $W_1^{(0)}(x)$ is solution of a quadratic equation which depends on a
polynomial $P^{(0)}(x)$ of degree $d-1$, i.e. $d$ variables remain to be fixed.

From the definition eq.\ref{defreso} of the correlation functions, considering the ${\cal A}_i$-cycles as circle, independent of $t$
around $\xi_i$\footnote{Indeed, when $t \to 0$, the cuts are reduced to double points at $\xi_i$'s. As $t$ grows,
these double points give rise to cuts of length of order ${n_i t\over n}$.}, one gets $d$ constraints
$$
\forall i=1, \dots, d \; , \; \;
{1 \over 2 i \pi} \oint_{{\cal A}_i} W_1^{(0)}(x) dx = {n_i t \over N}
$$
allowing to fix the coefficients of the polynomial $P^{(0)}(x)$, since the contour integral $\oint_{{\cal A}_i}$ pics only
one residue at $\xi_i$. One thus gets all the parameters of the spectral curve
as well as the one point function $W_1^{(0)}(x)$.

\smallskip

{\bf Properties of the one matrix model's spectral curve}

\medskip

The polynomial $H_{1MM}(x,y)$ has degree $2$ in $y$. This means that the embedding of ${\cal L}_{1MM}$
is composed by $2$ copies of the Riemann sphere glued by $g+1$ cuts so that the resulting Riemann surface ${\cal L}_{1MM}$ has genus $g$.
Each copy of the Riemann sphere corresponds to one particular branch of the solutions of the equation
$H_{1MM}(x,y)=0$.
Since there are only two sheets in involution, this spectral curve is said to be hyperelliptic. It also means that the application
$z \to \overline{z}$ is globally defined since it is the map which exchanges both sheets, i.e. which exchange the two
branches of the square root in \ref{W10}.

The Riemann surface ${\cal L}_{1MM}$ has genus $g$ lower than $d -1$\footnote{Notice that $d-1$ is an upper bound. It might happen that two branch points coincide resulting in the closing of one cut and decreasing of the genus by one. For some special value of the coefficients of the polynomial $H_{1MM}$ one can even get a genus zero spectral curve.
For application of matrix models to enumeration of surfaces, this very non-generic constraint is almost always satisfied (see \cite{eynform} for further considerations on this point).}.

The function $x(z)$ on the Riemann surface ${\cal L}_{1MM}$ has two simple poles (call them $\alpha_+$ and $\alpha_-$), one in each sheet.
Near $\alpha_\pm$, $y(z)$ behaves like:
$$
y(z) \mathop{\sim}_{z\to \alpha_+}   {t\over x(z)}  + O(1/x(z)^2)
$$
and
$$
y(z) \mathop{\sim}_{z\to \alpha_-} t_d x^d(z) + O(x^{d-1}(z)) .
$$

\subsection{Two point function}

Let us go one step further and consider the loop equation (\ref{loopeq}) for $k=2$ and $h=0$. It allows to obtain a formula
for $W_2^{(0)}(x,x_1)$:
$$
W_2^{(0)}(x,x_1)
= {{\partial \over \partial x_1}\,{W^{(0)}_{1}(x)-W^{(0)}_{1}(x_1)\over x-x_1} + P^{(0)}_{2}(x;x_1)\over 2 (V'(x)-W_1^{(0)}(x))}.
$$
A first look at this expression allows to see that this function is multivalued in term of the complex variable
$x$ and $x_1$. However, one can lift it to a monovalued function, actually a 2-form, on the spectral curve by
defining
$$
\widehat{\omega}_2^{(0)}(z,z_1):= W_2^{(0)}(x(z),x(z_1)) dx(z) dx(z_1).
$$
$\widehat{\omega}_2^{(0)}{\omega}(z,z_1)$ is thus a meromorphic bi-differential on ${\cal L}$. One can then
study all possible singularities of this formula and see that $\widehat{\omega}_2^{(0)}(z,z_1)$ has poles only
at $z \to \overline{z}_1$. On the other hand, the normalization of the two point function around the ${\cal A}$-cycles reads
$\oint_{{\cal A}_i} \widehat{\omega}_2^{(0)}(z,z_1) = 0$ for $i=1 , \dots, d$. These two conditions imply that
$\widehat{\omega}_2^{(0)}(z,z_1)$ is given by the Bergman kernel (see for example section 5.2.3 of \cite{EOrevue})
$$
\widehat{\omega}_2^{(0)}(z,z_1)= - B(z,\overline{z}_1) = B(z,z_1) - {dx(z) dx(z_1) \over (x(z)-x(z_1))^2}.
$$

\subsection{Correlation functions}

We have now everything in hand to compute any correlation function by solving the loop equations. First
of all, the study of the one and two point functions proved that it is more convenient to promote
the multivalued functions $W_n^{(h)}$ on the complex plane to monovalued meromorphic forms on ${\cal L}$\footnote{The
monovaluedness of the differential form $\omega_n^{(h)}$ on the spectral curve is obtained by induction on the Euler characteristic
$2h+n-2$ through the use of the loop equations (\ref{loopeq}).}:
$$
\omega_n^{(h)}(z_1,\dots,z_n):=
W_n^{(h)}(z_1,\dots,z_n) \prod_{i=1}^n dx(z_i) + \delta_{n,2} \delta_{h,0} {dx(z_1) dx(z_2) \over (x(z_1)-x(z_2))^2}
$$
and
$$ y(z) dx(z) := W_1^{(0)}(z) dz$$.
It is important to remember that the physical quantities encoded in the correlation functions are obtained as the terms of the expansion
of the latter when their variables approach the physical pole $\alpha_+$ of the spectral curve.

From the loop equations (\ref{loopeq}), one can prove by induction that $\omega_n^{(h)}(z_1,\dots,z_n)$ with $2h+n\geq 3$ can have
pole neither at coinciding points $x(z_i) = x(z_j)$, neither at the poles of $x$ nor at the double points. It may
have poles only at the branch points.

Let us now write down the Cauchy formula on the spectral curve:
$$
\omega_{n+1}^{(h)}(z,z_1,\dots,z_n) = \Res_{z' \to z} dS_{z',o}(z) \omega_{n+1}^{(h)}(z',z_1,\dots,z_n)
$$
where $o$ is an arbitrary point of ${\cal L}$. Since $\omega_{n+1}^{(h)}(z',z_1,\dots,z_n)$ has poles only
at the branch point $a_i$, moving the integration contours on ${\cal L}$ (and not $\mathbb{C}$!), one gets contributions from the
latter and the boundaries of the fundamental domain of ${\cal L}$  according to Riemann bilinear formula \cite{Farkas}
$$\begin{array}{l}
\omega_{n+1}^{(h)}(z,z_1,\dots,z_n) = - {\displaystyle \sum_i \Res_{z' \to a_i}} dS_{z',o}(z) \omega_{n+1}^{(h)}(z',z_1,\dots,z_n)
\cr
\hspace{3.6cm} + {\displaystyle \sum_{i=1}^g} \left[ \oint_{z' \in {\cal A}_i} B(z,z') \oint_{z' \in {\cal B}_i} \omega_{n+1}^{(h)}(z',z_1,\dots,z_n)\right. \cr
\hspace{4.4cm} \left. + \oint_{z' \in {\cal B}_i} B(z,z') \oint_{z' \in {\cal A}_i} \omega_{n+1}^{(h)}(z',z_1,\dots,z_n) \right] .\cr
\end{array}
$$
Since the correlation functions and the Bergmann kernel have vanishing ${\cal A}$-cycle integrals, the second and
third line vanish. One can then plug the expression for $\omega_{n+1}^{(h)}(z',z_1,\dots,z_n)$ coming from the loop
equations (\ref{loopeq}) into this formula. Since the polynomial $P_{n+1}^{(g)}(x(z'),z_1,\dots,z_n)$ is regular at the branch points,
it does not give any contribution and one gets the recursion formula
$$
\begin{array}{rcl}
\omega_{n+1}^{(h)}(z,z_1,\dots,z_n) &=& {\displaystyle \sum_i} \Res_{z' \to a_i} K(z,z')
\big[\omega_{n+2}^{(h-1)}(z',\overline{z'},z_1,\dots,z_n) \cr
&& +  {\displaystyle \sum_{j=0}^h \sum_{I \subset \{z_1,\dots,z_n\}}^{'}} \omega_{\left|I\right|+1}^{(j)}(z',I)\omega_{n-\left|I\right|+1}^{(h-j)}(\overline{z'},\left\{z_1,\dots,z_n\right\} \backslash I) \big] \cr
\end{array}
$$
where the sign $\sum^{'}$ means that the sum does not involve the terms with $(j,\left|I\right|) = (0,0)$ or
$(j,\left|I\right|) = (h,n)$ and the recursion kernel is
$$
K(z,z'):= {dS_{z',\overline{z'}}(z) \over 2 (y(z')-y(\overline{z'})) dx(z')}.
$$

It is easy to see that this recursive relation on $2h+n-2$, i.e. provided that $\omega_2^{(0)}(z,z_1) = B(z,z_1)$
is known, it determines all the other correlation functions through their topological expansion.

\br
This recursion can be graphically represented in such a way that it becomes very easy to remember and allows
to recover some of the properties of the correlation functions using only diagrammatic proofs. Details on this
diagrammatic representation can be found in \cite{EOrevue}.

\er

\subsection{Free energies}

In the preceding section, we have been able to compute the topological expansion of any correlation function $W_n$ for
$n>0$. Let us now address the case $n=0$, that is to say the computation of the topological expansion of the free energy.

For this purpose, one can build an operator acting from the space of $n+1$-differentials on ${\cal L}$ into
the space of $n$-differentials mapping the $n+1$-point function to the $n$-point function
\bt
For any $h$ and $n$ satisfying $2-2h-n<0$ and any primitive $\Phi$ of $ydx$, one has
$$
\omega_n^{(h)}(z_1,\dots,z_n) = {1 \over 2-2h-n} \sum_i \Res_{z \to a_i} \Phi(z) \omega_{n+1}^{(h)}(z,z_1,\dots,z_n).
$$
\et

One can guess that this definition can be extended to $n=0$ in order to get the topological expansion of the free
energies as follows:
\bt
The terms of the topological expansion of the free energy of the Hermitean one matrix model are given by:
$$
F^{(h)} = {1 \over 2-2h} \sum_i \Res_{z \to a_i} \Phi(z) \omega_1^{(h)}(z)
$$
for $h\geq 2$.
\et
This guess can be proved to be right by looking at the derivative of the result with respect to all the moduli of
the formal integral, i.e. the coefficient of the potential and the filling fractions. Indeed, they match the expected
variations of the free energies when varying these moduli \cite{ec1loopF}.

\sect{Matrices  coupled in a chain plus external field}\label{secChain}

It is remarkable that the recursive formula giving the topological expansion of the free energy and the
correlation functions depends on the moduli of the model only through the spectral curve. One can thus wonder wether the same
procedure can be applied to solve other matrix models which are known to be related to a spectral curve.
This is indeed the case for the model of two matrices coupled in chain \cite{CEO} but also for the
an arbitrary long chain of matrices in an external field \cite{Eynpratts}.

In order to deal with a large family of Hermitean matrix models at once, let us consider an arbitrarily long sequence
of matrices coupled in chain and submitted to the action of an external field.

The partition function is given by the chain of matrices formal matrix integral:
$$
Z_{\rm chain}=\int_{\rm formal} e^{-{N\over t}\Tr \left( {\displaystyle \sum_{k=1}^m} V_k(M_k) - {\displaystyle \sum_{k=1}^{m}} c_{k,k+1}\,M_k M_{k+1} \right)}\, dM_1\dots\, dM_m
$$
where the integral is a formal integral in the sense of the preceding section\footnote{The formal integral is
a power series in $t$ whose coefficients are Gaussian integrals. See \cite{eynform} for a review on this topic}, $M_{m+1}$ is a constant given diagonal matrix
$M_{m+1}=\Lambda$ with $s$ distinct eigenvalues $\lambda_i$ with multiplicities $l_i$:
$$
M_{m+1}=\Lambda= \hbox{diag}\left(\overbrace{\lambda_1,\dots,\lambda_1}^{l_1},\dots,\overbrace{\lambda_i,\dots,\lambda_i}^{l_i},\dots,\overbrace{\lambda_s,\dots,\lambda_s}^{l_s}\right)
$$
with ${\displaystyle \sum_i} l_i = N$ and one considers the $m$ polynomial potentials\footnote{It is possible to generalize all this section to potentials
whose derivative are arbitrary rational functions without any significant modification of the present procedure.} \\
$V_k(x) = -{\displaystyle \sum_{j=2}^{d_k+1}} {t_{k,j}\over j}\, x^j$.


As in the one matrix model case, the definition of the formal integral requires to choose around which saddle point one expands.
Saddle points are solutions of the set of equations
$$
\forall k=1,\dots,m,\qquad V'_k(\xi_k) = c_{k-1,k} \xi_{k-1}+c_{k,k+1}\xi_{k+1}
\qquad , \qquad
\exists j,\, \xi_{m+1} = \lambda_j
$$
which can be reduced to an algebraic equation with $D= s d_1 d_2 \dots d_m $ solutions.

This choice is thus equivalent to the choice of a $D$-partition $(n_1,\dots,n_D)$ of $N$ giving rise to the filling fractions:
$$
\epsilon_i = t {n_i \over N}
$$
for $i=1,\dots, D$ with $D= d_1 d_2 \dots d_m s$ and $n_i$ arbitrary integers satisfying
$$
\sum_i n_i = N.
$$

\subsubsection{Definition of the correlation functions}

The loop equations of the chain of matrices were derived in \cite{eynmultimat, Eynpratts}, and they
require the definition of several quantities.

For convenience, we introduce
$G_i(x_i):={1 \over x_i - M_i} = {\displaystyle \sum_{k=0}^\infty} {M_i^k \over x_i^{k+1}}$
as a formal power series in $x_i \to \infty$ as well as  a polynomial in $x$,
$
Q(x)= {1 \over c_{n,n+1}}\, {S(x) - S(\Lambda) \over x-\Lambda}
$,
where $S(x)$ is the minimal polynomial of $\Lambda$, $S(x)= {\displaystyle \prod_{i=1}^s} (x-\lambda_i)$.
We also define the polynomials $f_{i,j}(x_i,\dots,x_j)$ by $f_{i,j}=0$ if $j<i-1$, $f_{i,i-1}=1$, and
$$
f_{i,j}(x_i,\dots,x_j) = \det\left(\begin{matrix} V'_i(x_i) & -c_{i,i+1} x_{i+1} & & 0 \cr
-c_{i,i+1} x_i & V'_{i+1}(x_{i+1}) & \ddots & \cr
& \ddots & \ddots & -c_{j-1,j} x_j \cr
0 & & -c_{j-1,j} x_{j-1} & V'_j(x_j)
\end{matrix}\right)
$$
if $j\geq i$.
The latter satisfy the recursion relations
$$
c_{i-1,i} f_{i,j}(x_i,\dots,x_j) = V_i'(x_i) f_{i+1,j}(x_{i+1},\dots,x_j) - c_{i,i+1} \, x_i \, x_{i+1} \, f_{i+2}(x_{i+2},\dots,x_j).
$$

Let us finally define the correlation functions and some useful auxiliary functions. In the following
$\hbox{Pol}_{x} f(x)$ refers to the polynomial part of $f(x)$ as $x \to \infty$.
For $i=2,\dots , m$, we define
$$
\begin{array}{l}
W_i(x_1,x_i,\dots,x_m,z):= \cr
{\displaystyle \hbox{Pol}_{x_i,\dots,x_m}} f_{i,m}(x_i,\dots,x_m) \left< \Tr \left( G_1(x_1) G_i(x_i) \dots G_m(x_m)Q(z)\right)\right>,\cr
\end{array}
$$
which is a polynomial in variables $x_i,\dots,x_m,z$, but not in $x_1$,
for $i=1$,
$$
\begin{array}{l}
W_1(x_1,x_2,\dots,x_m,z):= \cr
 \hbox{Pol}_{x_1,\dots,x_m} f_{1,m}(x_1,\dots,x_m) \left< \Tr \left( G_1(x_1) G_2(x_2) \dots G_m(x_m)Q(z)\right)\right> \cr
\end{array}
$$
which is a polynomial in all variables and, for $i=0$, $W_0(x) = \left< \Tr G_1(x) \right>$.
We also define:
$$
\begin{array}{l}
 W_{i;1}(x_1,x_i,\dots,x_m,z;x_1') :=\cr
 {\displaystyle \hbox{Pol}_{x_i,\dots,x_m}} f_{i,m}(x_i,\dots,x_m) \left< \Tr \left(G_1(x_1')\right) \Tr \left( G_1(x_1) G_i(x_i) \dots G_m(x_m)Q(z)\right)\right>_c . \cr
\end{array}
$$
All these functions admit a topological expansion:
$$
W_i = \sum_g (N/t)^{1-2g} W_i^{(g)}
\hspace{0.8cm} \hbox{and} \hspace{0.8cm}
W_{i;1} = \sum_g (N/t)^{-2g} W_{i;1}^{(g)}.
$$

\subsubsection{Loop equations and spectral curve}

In this model, the master loop equation reads \cite{eynmultimat, Eynpratts}:
\begin{equation}\label{masterloopchain1}
\begin{array}{l}
W_{2;1}(x_1,\dots ,x_{m+1};x_1) + {t\over N} W_1(x_1, \dots, x_{m+1})
- \left( V_1'(x_1) - c_{1,2} x_2 \right) S(x_{m+1})\cr
+ (c_{1,2} x_2 - V_1'(x_1) +{t\over N}W_0(x_1))
\Big( {t\over N}W_2(x_1, \dots ,x_{m+1}) - S(x_{m+1})\Big)  \cr
=  {t\over N} {\displaystyle \sum_{i=2}^m}
\left(V_i'(x_i)-c_{i-1,i} x_{i-1} - c_{i,i+1} x_{i+1}\right) W_{i+1}(x_1,x_i,\dots,x_{m+1}). \cr
\end{array}
\end{equation}
Let us consider specific values for the variables $x_i$ in order to turn it into an equation involving only $x_1$ and $x_2$.
One defines $\left\{\hat x_i(x_1,x_2)\right\}_{i=3}^{m+1}$ as functions of the two first variables $x_1$ and $x_2$ by
\begin{equation}\label{recxichainmat}
c_{i,i+1} \hat{x}_{i+1}(x_1,x_2) = V_i'(\hat{x}_i(x_1,x_2)) - c_{i-1,i} \hat{x}_{i-1}(x_1,x_2) .
\end{equation}
for $i=2,\dots,m$ with the initial conditions $\hat x_1(x_1,x_2)=x_1$ and $\hat x_2(x_1,x_2)=x_2$.

Choosing $x_i=\hat x_i(x_1,x_2)$, reduces the master loop equation to an equation in $x_1$ and $x_2$:
$$
 \widehat{W}_{2;1}(x_1,x_2;x_1) + {t\over N}\left(c_{1,2} x_2 - Y(x_1)\right) \widehat{U}(x_1,x_2) = \,\widehat{E}(x_1,x_2)
$$
where $Y(x) = V_1'(x) - {t\over N}W_0(x)$, the hat means that the functions are considered at the value $x_i = \hat{x}_i(x_1,x_2)$,
i.e. $\widehat(f)(x_1,x_2):=f(x_1,x_2,\hat{x}_3,\hat{x}_4,\dots,\hat{x}_n)$ for an arbitrary function $f$, and one has defined
$$
\widehat{U}(x_1,x_2) = W_2(x_1,x_2,\hat{x}_3,\dots,\hat{x}_{m+1}) - {N\over t}S(\hat{x}_{m+1}),
$$
and
$$
\widehat{E}(x_1,x_2) = - {t\over N}\widehat{W}_1(x_1,x_2) + \left(V_1'(x_1)-c_{1,2} x_2\right) \widehat{S}(x_1,x_2).
$$

Finally, the leading order in the topological expansion gives
\begin{equation}\label{masterloopleading}
\widehat{E}^{(0)}(x_1,x_2) = \left(c_{1,2} x_2 - Y^{(0)}(x_1)\right) \widehat{U}^{(0)}(x_1,x_2)
\end{equation}
where one should notice that $\widehat{W}_1(x_1,x_2)$, and thus $\widehat{E}(x_1,x_2)$, is a polynomial in both $x_1$ and $x_2$.

Again, this equation is valid for any $x_1$ and $x_2$, and, if we choose $x_2$ such that $c_{1,2} x_2 = Y^{(0)}(x_1)$, we get:
\begin{equation}\label{spcurvechainmat}
H_{chain}(x_1,x_2) := \widehat{E}^{(0)}(x_1,x_2) = 0.
\end{equation}
This algebraic equation is the spectral curve of our model.

\bigskip

{\bf Study of the spectral curve}

The algebraic plane curve $H_{chain}(x_1,x_2)=0$, can be parameterized by a variable $z$ living on a compact Riemann
surface ${\cal L}_{chain}$ of some genus $g$, and two meromorphic functions $x_1(z)$ and $x_2(z)$ on it. Let us study
it in greater details.

The polynomial $H_{chain}(x_1,x_2)$ has degree $1+{D\over d_1} $ (resp. $d_1+ {D\over d_1 d_2}$)  in $x_2$ (resp. $x_1$).
This means that the embedding of ${\cal L}_{chain}$ is composed by $1+{D\over d_1}$ (resp. $d_1+{D\over d_1d_2}$)
copies of the Riemann sphere, called $x_1$-sheets (resp. $x_2$-sheets), glued by cuts so that the resulting Riemann surface ${\cal L}_{chain}$
has genus $g$. Each copy of the Riemann sphere corresponds to one particular branch of the solutions of the equation
$H_{chain}(x_1,x_2)=0$ in $x_2$ (resp. $x_1$).

The Riemann surface ${\cal L}_{chain}$ has genus $g$ lower than $D -s$
with $D=s\,d_1\dots d_m$.

One can consider all the variables $x_i(p):=\hat{x}_i(x_1(p),x_2(p))$ as meromorphic functions on ${\cal L}_{chain}$ as opposed to
only $x$ and $y$ in the one matrix model case. Their
negative divisors are given by
$$
\left[x_k(p)\right]_- = - r_k \infty - s_k \sum_{i=1}^s \hat{\lambda}_i
$$
where $\infty$ is the only point of ${\cal L}_{chain}$ where $x_1$ has a simple pole, the $\hat{\lambda}_i$ are the preimages
of $\lambda_i$ under the map $x_{m+1}(p)$,
$x_{m+1}(\hat{\lambda}_i) = \lambda_i$,
and the degrees $r_k$ and $s_k$ are integers given by
$r_1:=1$, $r_k:= d_1 d_2 \dots d_{k-1}$, $s_{m+1}:=0$, $s_m:=1$, and $s_k:= d_{k+1} d_{k+2} \dots d_m\, s$.

Note that the presence of an external matrix creates as many poles as the number of distinct eigenvalues of this external
matrix $M_{m+1}=\Lambda$\footnote{The cases of matrix models without external field correspond to a totally degenerate external matrix
$\Lambda= c\, {\rm Id}$ with only 1-eigenvalue. There are thus two poles as in the 1 matrix model studied earlier.}.

\subsubsection{Solution of the loop equations}

The procedure used to solve the loop equations in the one matrix model cannot be generalized in this setup,
mainly because the involution $z \to \overline{z}$ is not globally defined on the spectral curve.
However, the loop equations can been solved
using a detour \cite{Eynpratts}.
This resolution proceeds in three steps.
One first shows that the loop equations eq.\ref{masterloopchain1} have a unique solution admitting a topological expansion.
One then finds
a solution of these equations:
\bl\label{completecurvechain}
\begin{equation}\label{lemspcurve}
E(x(z),y) = -K\,"\left<\prod_{i=0}^{d_2} (y-V'_1(x(z^i))+{t \over N} \Tr{1\over x(z^i)-M_1}) \right>_c"
\end{equation}
where $K$ is a constant and the inverted comas $"<.>"$ means that every time one encounters a two point
function $\left<\Tr{1\over x(z^i)-M_1}\Tr{1\over x(z^j)-M_1}\right>_c$, one replaces it by
$W_{0;1}^{(0)}(z_i;z_j):=\left<\Tr{1\over x(z^i)-M_1}\Tr{1\over x(z^j)-M_1}\right>_c + {1 \over (x(z_i)-x(z_j))^2}$.
\el
The matching of the coefficients of the polynomials in $y$ in the left- and right hand sides of \ref{lemspcurve} and
a few algebro-geometrical
computations allows to solve the loop equations\footnote{See \cite{CEO,Eynpratts} for the detailed proof.} to get
\bt\label{thFgchainmat}
The correlation functions of a chain of matrices formal integral are recursively obtained by computing residues on ${\cal{L}}_{chain}$:
\begin{equation}\label{defWG}
\begin{array}{rcl}
\omega_{n+1}^{(h)}(z,z_1,\dots,z_n) &=& {\displaystyle \sum_i} \Res_{z' \to a_i} K(z,z')
\big[\omega_{n+2}^{(h-1)}(z',\overline{z'},z_1,\dots,z_n) \cr
&& +  {\displaystyle \sum_{j=0}^h \sum_{I \subset \{z_1,\dots,z_n\}}^{'}} \omega_{\left|I\right|+1}^{(j)}(z',I)\omega_{n-\left|I\right|+1}^{(h-j)}(\overline{z'},\left\{z_1,\dots,z_n\right\} \backslash I) \big] \cr
\end{array}
\end{equation}
where, as in the preceding section,
\begin{eqnarray*}
\omega_n^{(h)}(z_1,\dots,z_n)
&=& \Res_{N \to \infty} N^{n+2h-3} \left< \prod_{i = 1}^n \Tr G_1(x_i)
\right>_c\,dx(z_1)\dots dx(z_n) \cr
&&  + \delta_{n,2}\delta_{g,0}\,{dx(z_1)dx(z_2)\over (x(z_1)-x(z_2))^2} .
\end{eqnarray*}
the two point function $\omega_2^{(0)}$ is the Bergman kernel of the spectral curve ${\cal L}_{chain}$, the recursion kernel is
$$
K(z,z'):= {dS_{z',\overline{z'}}(z) \over 2 (y(z')-y(\overline{z'})) dx(z')},
$$
$x$ and $y$ are two meromorphic functions on ${\cal{L}}_{chain}$ such that $H_{chain}(x(z),y(z))=0$ for any point
$z \in {\cal {L}}_{chain}$ and $a_i$ are the $x$-branch points, i.e. solutions to $dx(a_i) = 0$.

\et
Remember that $H_{chain}$ was defined in \ref{spcurvechainmat} by $H_{chain}(x_1,c_{12} x_2)=0$ for $c_{12} x_2 = Y^{(0)}(x_1)$.
Thus the function $x$ and $y$ can be thought of as continuation to the whole spectral curve of $x_1$ and $c_{12} x_2$
respectively.

The free energy is also obtained by using the same formula as in the one matrix case:
\bt
For $h>1$ and any primitive $\Phi$ of $y \, dx$, one has
\begin{equation}\label{defF}
F^{(h)} = {1 \over 2-2h} \sum_i \Res_{z \to a_i} \Phi(z) \omega_1^{(h)}(z).
\end{equation}
\et

Thus, the solution of any chain-matrix model with an external field is obtained by the exact same formula as the
solution of the one matrix model: the only difference is the spectral curve used to apply this recursion.

\sect{Generalization: topological recursion}\label{conclusion}

We have seen that the loop equation method gives a unique solution for a large family of formal matrix models.
The only input of this solution is the spectral curve of the considered model. In \cite{EOinvariants}, it has been
proposed to use equations \ref{defWG} and \ref{defF}
to associate infinite sets of correlation functions and free energies to
any spectral curve $({\cal L},x,y)$ where ${\cal L}$ is a compact Riemann surface and $x$ and
$y$ two functions analytic in some open domain of ${\cal L}$.

The free energies and correlation functions built from this recursive procedure show many interesting properties
such as invariance under a large set of transformations of the spectral curve, special geometry relations,
modular invariance or integrable properties. In particular, it is a very convenient tool to study critical
regimes and get the universal properties of the matrix integrals described in the chapter 6 of this volume.
It is also very useful to compare different matrix integrals. Eventually, this procedure proved to be efficient
in the resolution of many problems of enumerative geometry or statistical physics such as string theory,
Gromov-Witten invariants theory, Hurwitz theory or exclusion processes such as TASEP or PASEP. Most of
the results proposed by this approach are still conjectures up to now but the numerous checks passed so far
tends to prove that this generalization of the loop equation method is a very promising field\footnote{For a review
on this subject, see \cite{EOrevue} and references therein.}.

The inductive procedure presented in this chapter only allows to compute one particular set of observables of
multi-matrix models. It does not compute correlation functions involving more than one type of matrix inside the
same trace. These more complicated objects are very important for their application to quantum gravity or conformal
field theories where they correspond to the insertion of boundary operators. In the two matrix model, the loop equation
method allowed to compute the topological expansion of any of these operators \cite{EOallmixed}. In the chain of matrices case,
only a few of them were computed in their large $N$ limit only \cite{eynmultimat}. The computation of any observable of
the chain of matrices is still an open problem which is very likely to be solved by the use of loop equations.

\ \\
{\sc Acknowledgements}:
It is a pleasure to thank Bertrand Eynard who developed most of the material exposed in this chapter and patiently
taught me all I know about these topics.

\end{document}